\documentclass{article}
\usepackage{times}
\begin{document}
\title{Quantum--Mechanical Dualities on the Torus}
\author{Jos\'e M. Isidro\\
Instituto de F\'{\i}sica Corpuscular (CSIC--UVEG)\\
Apartado de Correos 22085, Valencia 46071, Spain\\
{\tt jmisidro@ific.uv.es}}

\maketitle

\begin{abstract}

On classical phase spaces admitting just one complex--differentiable structure, 
there is no indeterminacy in the choice of the creation operators that create quanta out of a given vacuum. 
In these cases the notion of a quantum is universal, {\it i.e.}, independent of the observer on classical phase space. 
Such is the case in all standard applications of quantum mechanics. However, recent developments suggest that the notion 
of a quantum may not be universal. Transformations between observers that do not agree on the notion 
of an elementary quantum are called dualities. Classical phase spaces admitting more than one 
complex--differentiable structure thus provide a natural framework to study dualities in quantum mechanics. 
As an example we quantise a classical mechanics whose phase space is a torus and prove explicitly 
that it exhibits dualities.


2001 Pacs codes: 03.65.Bz, 03.65.Ca, 03.65.-w.

\end{abstract}

\tableofcontents

\section{Introduction}\label{mariconestodos}

In the quantum mechanics of a finite number of degrees of freedom, 
a {\it duality}\/ is the possibility of having nontrivial transformations between the
Hilbert spaces of states corresponding to different observers on classical
phase space ${\cal C}$ \cite{VAFA}.  Under an {\it observer}\/ one
understands, in general relativity, a little man carrying a ruler and a clock. 
In fact one may forget about the little man while keeping his ruler and his clock, 
to conclude that an observer is just a local coordinate chart on spacetime.
By the same token, in a quantum--mechanical setup, an observer will be a
local coordinate chart $({\cal U}, z)$ on ${\cal C}$, where ${\cal U}$ is 
an open neighbourhood of a point $p\in{\cal C}$ endowed with the coordinate
functions $z$.  

The notation $z$ for the coordinates on ${\cal C}$ is not accidental. As a
classical phase space, ${\cal C}$ will be a symplectic manifold. As the
starting point for quantisation, ${\cal C}$ will also be a complex manifold,
{\it i.e.}, coordinate changes $({\cal U},z)\rightarrow ({\cal U}', z')$ will be 
biholomorphic maps between the charts ${\cal U}$ and ${\cal U}'$. This is 
so because, in the quantum theory, provision has to be made for the
creation and annihilation of quanta. The quantum--mechanical creation and 
annihilation operators are easily seen to be intimately
related with the complex structure on ${\cal C}$. In other
words, quantisation requires a complex--differentiable structure on classical phase space.
It is natural to assume that the complex structure and the symplectic structure
are compatible, {\it i.e.}, that Darboux coordinates $(q,p)$ on ${\cal C}$ are the
real and imaginary parts of the holomoprhic coordinates $z$, so we can
write $z=q+{\rm i}p$. However it must be realised that, in classical
mechanics, using complex coordinates $z$ or real Darboux coordinates $(q,p)$ 
is immaterial, as the relevant object to consider is the algebra of smooth
functions on ${\cal C}$.  It is only upon quantisation that a complex structure
becomes relevant. This fact becomes especially clear in coherent--state
quantisation. We will see that also canonical quantisation can be regarded
as the choice of a complex structure on ${\cal C}$. (Symplectic spaces
admitting no complex structure, but just an almost complex structure, have
been analysed in ref. \cite{MODP}; all symplectic spaces are almost
complex).

On classical phase spaces admitting just one complex structure, there is
no indeterminacy in the choice of the creation operators that create quanta out of 
a given vacuum. In these cases the notion of a quantum is universal, {\it i.e.}, 
independent of the observer on classical phase space. Such is the case in all standard applications 
of quantum mechanics (harmonic oscillator, central potentials, angular momentum). 
However, as pointed out in ref. \cite{VAFA}, recent developments suggest
considering the possibility that the notion of a quantum may not 
be universal. Classical phase
spaces admitting more than one complex structure thus provide a natural
framework to study dualities in quantum mechanics; this is what we do in
this article. In particular, we
consider a classical mechanics whose phase space is a torus $T^2$ (a compact Riemann surface
with genus 1), which is the simplest case of a complex manifold admitting 
a nontrivial moduli space of nonbiholomorphic complex structures. 
We are then faced with the unusual situation that not only 
the coordinate $q$, but also its conjugate momentum $p$, is a compact
variable. Moreover, there is a whole continuum of nonbiholomoprhic, hence
nonequivalent, choices for the complex structure $\tau$ on $T^2$
\cite{SCHLICHENMAIER}. Each one of them defines a physically different
quantum theory on $T^2$; dualities arise as nonbiholomorphic transformations 
between them.

This article is organised as follows. Section \ref{labastidacabron} presents a
specific classical dynamics whose classical phase space is a torus.
Despite its unusual aspect, this dynamics actually turns out to be a
natural generalisation of well--known examples, such the harmonic
oscillator and the sine--Gordon model. 
The unifying character of our model appears when different limits are
taken: then it reproduces the above--mentioned theories in a natural way.
In section \ref{pajarescasposo} we quantise it canonically. We perform a detailed
analysis of the Hilbert space of states, the operators acting on it, 
and especially of the vacuum state, to arrive at the conclusion that
the complex structure of the torus is a key ingredient in the quantum
theory. In this way we conclude that our model exhibits dualities in the 
sense of refs. \cite{MODP, PQM}. For the sake of simplicity 
we work with just one degree of freedom, but our conclusions can be easily generalised.
Thus in section \ref{alvarezgaumequetepartaunrayo} 
we provide more examples of classical phase spaces that are modelled on the torus.
Section \ref{muchasoberbia} discusses our results. 

Issues partially overlapping with ours are dealt with in refs. \cite{MATONE, MINIC}. Finally, the (in)dependence of the quantum theory 
on the complex structure has been analysed in ref. \cite{WITTEN}; see also \cite{WOODHOUSE}. However our perspective is  entirely 
different, since we interpret this dependence as a duality, a notion that chronologically appeared later \cite{VAFA}.

\section{Classical mechanics on the torus}\label{labastidacabron}

Let us consider the $(q,p)$ plane. Quotienting the latter by the periodicity $q-\pi \simeq q+\pi$ 
we obtain an infinitely tall cylinder $Y^2$. A torus $T^2$ is the quotient of $Y^2$
obtained by imposing the additional periodicity $p-\pi\simeq p+\pi$.
A classical mechanics whose phase space is this $T^2$ is given by the Hamiltonian function
\begin{equation}
H = -\cos p - \cos q,
\label{labastidahijoputa}
\end{equation}
where both $p$ and $q$ run over the interval $(-\pi, \pi)$. 
More precisely, above we have expressed the Hamiltonian in one particular coordinate 
chart on $T^2=S^1\times S^1$. Now $S^1$ cannot be covered by just one coordinate 
$\varphi$ running over $(-\pi, \pi)$. However it can be covered by two charts, 
{\it e.g.}, one with the coordinate $\varphi_1\in (-\pi, \pi)$, the other with 
the coordinate $\varphi_2\in (0, 2\pi)$. In this way we obtain an atlas 
on $T^2$ containing 4 charts. In this section we will work on the chart specified 
in eqn. (\ref{labastidahijoputa}). 

Separating out the quadratic term in $\cos p = \sum_k(-1)^kp^{2k}/(2k)!$, we can interpret $H$ as the dynamics 
of a unit--mass particle with a $p$--dependent potential $V(q,p)$:
\begin{equation}
H={1\over 2} p^2 + V(q,p), \qquad V(q,p)=-\cos q 
-\sum_{k=2}^{\infty}{(-1)^k\over (2k)!} p^{2k}.
\label{ppexn}
\end{equation} 
Neglecting terms in $p$ higher than quadratic one obtains the
sine--Gordon Hamiltonian
\begin{equation}
H_{\rm SG}=\frac{1}{2}p^2 -\cos q,
\label{cojonudo}
\end{equation}
which describes a mathematical pendulum within a constant gravitational
field. Further neglecting terms in $q$ higher than quadratic corresponds
to taking the limit of small oscillations, given by the harmonic oscillator
\begin{equation}
H_{2}= {1\over 2} \left( p^2 + q^2\right).
\label{exqp}
\end{equation}
Eqn. (\ref{cojonudo}) may  be regarded as the approximation to $T^2$ given by the
cylinder $Y^2$, where $q$ remains periodic while $p$ becomes noncompact.
Similarly, eqn. (\ref{exqp}) may be understood as the tangent--space approximation to
$T^2$, given by the $(q,p)$ plane, where both $q$ and $p$ are noncompact.
Conversely, the Hamiltonian (\ref{labastidahijoputa}) can be regarded as a large--$p$,
large--$q$ completion of the approximations (\ref{cojonudo}), (\ref{exqp}).

\section{Quantisation}\label{pajarescasposo}

Let us now set $z=q+{\rm i}p$. This defines a complex structure on the $(q,p)$ plane.
In coherent--state quantisation \cite{PERELOMOV}, complex analyticity is manifest all along,
through the use of the variable $z$. However, since our quantisation will be canonical, 
it will be convenient to retain $q,p$ as our variables. We will lose manifest analyticity, 
but we will nevertheless realise that the quantum theory depends crucially on the underlying complex structure.

\subsection{The Hilbert space}\label{ramallocasposo}

In canonical quantisation there is one quantum state per unit of symplectic volume on ${\cal C}$. 
Thus a compact, $2N$--dimensional phase space gives rise to a finite--dimensional Hilbert space. 
Moreover, the dimension of ${\cal H}$ is a monotonically increasing function of the symplectic volume 
of ${\cal C}$. We express this functional relation as
\begin{equation}
\int_{\cal C}\omega^N \sim {\rm dim}\, {\cal H},
\label{gguu}
\end{equation}
where $\omega$ is the symplectic 2--form on ${\cal C}$, and the sign $\sim$ 
means that we suppress normalisation factors such as $2\pi$, ${\rm i}\hbar$, etc.
Unless the $2N$--form $\omega^{N}$ is of fast decrease at infinity, a noncompact ${\cal C}$ 
leads to an infinite--dimensional Hilbert space ${\cal H}$.

The Hamiltonian (\ref{labastidahijoputa}) refers to a specific complex structure, namely, a square torus of sides $2\pi$ 
and local holomorphic coordinate $z=q+{\rm i}p$. An arbitrary complex structure on $T^2$ would correspond to the Hamiltonian
\begin{equation}
H=-\cos \alpha p - \cos (\beta q +\delta),
\label{labastidaputamierda}
\end{equation}
where $\alpha,\beta\in{\bf R}$ are arbitrary nonzero parameters allowing for different lengths along the torus axes, 
and $\delta\in(0,2\pi)$ is an arbitrary phase difference between the two periodicities. Setting $\alpha=1$, 
the parameter $\beta$ becomes the ratio between those lengths. Thus the remaining two parameters 
$0<\beta<\infty$, $0<\delta<2\pi$ respectively become the modulus and the argument of
the modular parameter $\tau$ in the geometry of complex tori \cite{SCHLICHENMAIER}.
A local holomorphic coordinate on this torus is $w=(\beta q + \delta) + {\rm i}p$.
The corresponding Hamiltonian, that we will work with from now on, is 
\begin{equation}
H=-\cos p - \cos (\beta q +\delta).
\label{labastidatecalzounahostia}
\end{equation}
The complex structure of (\ref{labastidatecalzounahostia}) carries a symplectic form $\omega_{\beta,\delta}$ 
\begin{equation}
\omega_{\beta,\delta}={\rm d} p\wedge{\rm d}(\beta q +
\delta)=\beta\,{\rm d}p\wedge{\rm d}q.
\label{ffmm}
\end{equation}
The coordinate transformation 
\begin{equation}
z=q+{\rm i}p\longrightarrow w=(\beta q+\delta)+{\rm i}p
\label{gatita}
\end{equation}
is canonical for $\beta=1$ and for any $\delta\in(0,2\pi)$.
In the spirit of the $\sim$ sign in eqn. (\ref{gguu}), we can assume $\omega_{\beta=1,\delta}$ normalised as
\begin{equation}
\int_{T^2}\omega_{1,\delta} = 1.
\label{nsvu}
\end{equation}
Eqn. (\ref{nsvu}) corresponds to a 1--dimensional Hilbert space. If $n$ denotes the whole part of $\beta\geq 1$, the symplectic form $\omega_{\beta,\delta}$ corresponds to an $n$--dimensional Hilbert space.

\subsection{Position and momentum}\label{labacasposo}

In coordinate representation, a basis of ${\cal H}={\bf C}^n$ is given by a set of $n$ orthonormal states $\vert q_j\rangle$:
\begin{equation}
\sum_{j=1}^n \vert q_j\rangle\langle q_j\vert = {\bf 1}.
\label{labastidamierda}
\end{equation}
The position operator $Q$ is defined as
\begin{equation}
Q\vert q_j\rangle = q_j \vert q_j\rangle,\qquad j=1,\ldots, n.
\label{alvarezgaumecabron}
\end{equation}
We define an operator $U$
\begin{eqnarray}
U\vert q_j\rangle&=&-{\rm i}\vert q_{j+1}\rangle, \quad j=1, \ldots
n-1,\nonumber\\
U\vert q_n\rangle&=&-{\rm i}\vert q_1\rangle.
\label{cagonlabastida}
\end{eqnarray}
Their commutator is
\begin{equation}
\left[Q, U\right]\vert q_j\rangle=-{\rm i}(q_{j+1}-q_{j})\vert
q_{j+1}\rangle,\qquad j=1,\ldots,n.
\label{muchamierda}
\end{equation}
The above holds for $j=n$ under the identification $n+1\simeq 1$, {\it i.e.},
\begin{equation}
q_{n+1}=q_1,\qquad\vert q_{n+1}\rangle=\vert q_1\rangle.
\label{kakagomez}
\end{equation}
While $Q$ is selfadjoint, $U$ is unitary:
\begin{equation}
U^{+}\vert q_j\rangle={\rm i}\vert q_{j-1}\rangle =
U^{-1}\vert q_j\rangle,\qquad j=1,\ldots,n.
\label{mierdacesargomez}
\end{equation}
This is no surprise, as $U$ effects a finite shift on the $\vert q_j\rangle$. 
As such it must equal the exponential of ($\sqrt{-1}$ times) some selfadjoint momentum 
operator $P$. However the latter has no action defined on the $\vert q_j\rangle$ which, 
by their very nature, are discrete. This has profound consequences. A
selfadjoint momentum operator $P$ could be defined as
\begin{equation}
P=-{\rm i}\log U,
\label{cagonramallo}
\end{equation}
and its action on the $\vert q_j\rangle$ would then be some linear combination
\begin{equation}
P\vert q_j\rangle = \sum_{k=1}^np_{jk}\vert q_k\rangle,\qquad j=1,\ldots,n.
\label{shitexists}
\end{equation}
One could now try and impose the Heisenberg algebra
\begin{equation}
\left[Q,P\right]={\rm i},
\label{twerq}
\end{equation}
in order to obtain the matrix $p_{jk}$ of eqn. (\ref{shitexists}).
However, on $T^2$ there is no way we can satisfy the commutator
(\ref{twerq}): this would violate the Stone--von Neumann theorem
because our Hilbert space is ${\bf C}^n$ \cite{NEUMANN}. In practical terms, 
some simple algebra shows that the $p_{jk}$ in fact cannot be obtained
by imposing the above commutator. The closest analogue of the Heisenberg
algebra that we can get in our model is eqn. (\ref{muchamierda}).

\subsection{The Hamiltonian}\label{alvarezgaumecasposo}

The quantum Hamiltonian corresponding to the classical Hamiltonian
(\ref{labastidatecalzounahostia}),
\begin{equation}
H=-\cos  P-\cos (\beta Q+\delta),
\label{gomezcasposo}
\end{equation}
can be expressed as
\begin{equation}
H=-\frac{1}{2}\left({\rm e}^{{\rm i} P}+{\rm e}^{-{\rm i} P}\right)-\cos 
(\beta Q+\delta)
=-\frac{1}{2}\left(U+U^{+}\right)-\cos (\beta Q+\delta).
\label{mascasposo}
\end{equation}
It acts on the basis $\vert q_j\rangle$ as
\begin{equation}
H\vert q_k\rangle = \frac{{\rm i}}{2}\left(\vert q_{k+1}\rangle-\vert
q_{k-1}\rangle\right) - \cos (\beta q_k + \delta) \vert q_k\rangle,
\label{mariconmas}
\end{equation}
with the periodicity (\ref{kakagomez}) understood. Hence the Hamiltonian
matrix
\begin{equation}
\langle q_j\vert H\vert q_k\rangle = \frac{{\rm i}}{2}\left(\delta_{j,k+1}-\delta_{j,k-1}\right) -\delta_{jk}\,\cos 
(\beta q_j + \delta).
\label{gomezcabron}
\end{equation}
We will compute the characteristic polynomial $s_n(E)={\rm det}\,(E{\bf 1}-H)$ in 
the energy $E$ for an arbitrary $n$, but let us first work out some examples. 
When $n=3$ we have
\begin{equation}
\prod_{j=1}^3\left(\cos (\beta q_j + \delta) +E\right)+\left(\frac{{\rm i}}{2}\right)^2\sum_{j=1}^3\left(\cos (\beta q_j +
\delta)+E\right).
\label{barbonladron}
\end{equation}
When $n=4$ the characteristic polynomial is
$$
\prod_{j=1}^4\left(\cos (\beta q_j + \delta)+E\right)+\left(\frac{{\rm i}}{2}\right)^2
\sum_{(k,l)\atop{\rm cyclic}}\left(\cos (\beta q_k + \delta)+E\right)\left(\cos 
(\beta q_l + \delta)+E\right),
$$
where the subindex {\it cyclic}\/ in the summation means that one is 
to sum over the pairs 
$$
(k,l)=(1,2)(2,3)(3,4)(4,1).
$$
When $n=5$ we have a characteristic polynomial
$$
\prod_{j=1}^5\left(\cos (\beta q_j + \delta)+E\right)
+\left(\frac{{\rm i}}{2}\right)^4\sum_{j=1}^5\left(\cos (\beta q_j +
\delta )+E\right)
$$
$$
+\left(\frac{{\rm i}}{2}\right)^2\sum_{(j,k,l)\atop{\rm
cyclic}}\left(\cos(\beta q_j + \delta)+E\right)\left(\cos (\beta q_k +
\delta) +E\right)\left(\cos (\beta q_l +\delta )+E\right),
$$
where the subindex {\it cyclic}\/ in the summation means that one is 
to sum over the triples 
$$
(j,k,l)=(1,2,3)(2,3,4)(3,4,5)(4,5,1)(5,1,2).
$$
One proves by induction that, for even $n$, the characteristic polynomial is  
\begin{equation}
{\rm det}\,\left(E{\bf 1}-H\right)=
\sum_{p=2\atop{\rm even}}^n\left(\frac{{\rm
i}}{2}\right)^{n-p}\sum_{(l_1,l_2,\ldots,l_p)\atop {\rm
cyclic}}\prod_{k=1}^p\left(\cos (\beta q_{l_k} + \delta)+E\right),
\label{genial}
\end{equation}
while for odd $n$ it is
\begin{equation}
{\rm det}\,\left(E{\bf 1}-H\right)=
\sum_{p=1\atop{\rm odd}}^n\left(\frac{{\rm
i}}{2}\right)^{n-p}\sum_{(l_1,l_2,\ldots,l_p)\atop {\rm
cyclic}}\prod_{k=1}^p\left(\cos(\beta q_{l_k} + \delta)+E\right).
\label{esplendido}
\end{equation}
The subindex {\it cyclic}\/ means that the corresponding sum extends to all $p$--tuples
$$
(l_1,l_2\ldots,l_p)=(1,2,\ldots, p)(2,3,\ldots, p+1)(3,4,\ldots p+2)\ldots
$$
$$
\ldots (n-p+1,n-p+2\ldots,n)(n-p+2,n-p+3,\ldots, 1)
(n-p+3,n-p+4,\ldots,2)\ldots
$$
\begin{equation}
\ldots (n,1,\ldots, p-2,p-1).
\label{magnifique}
\end{equation}
The cyclic property is a consequence of eqn. (\ref{kakagomez}). A term
common to both (\ref{genial}) and (\ref{esplendido}) is
$$
\prod_{j=1}^n\left(\cos (\beta q_j + \delta)+E\right),
$$ 
as it corresponds to the unique $n$--tuple $(1,2,\ldots, n-1,n)$. All factors of $\sqrt{-1}$ 
are raised to even powers, so the characteristic polynomial has real coefficients.  That the roots are 
also real follows from the Hermitian property of $H$ in eqn. (\ref{gomezcabron}).

\subsection{The vacuum}\label{pajarescabron}

We claim that the vacuum $\vert 0\rangle$ of eqn. (\ref{gomezcabron}) is nondegenerate. To simplify the argument we will 
first set $\beta=1$, $\delta = 0$. In order to prove nondegeneracy, we first assume turning off the kinetic term $-\cos P$ 
in the Hamiltonian (\ref{gomezcasposo}). Generically the spectrum of $-\cos Q$ is 2--fold degenerate because the cosine 
function is even, $\cos x = \cos(-x)$. The minimum of the potential is attained by $q=0$; at this point the parity transformation 
$x\to -x$ does not lead to a 2--fold degeneracy.  Since the kinetic term has been switched off we can assume, 
without loss of generality, that the vacuum $\vert 0\rangle$ is the state $\vert q_1\rangle$. Then the pair of states 
$\vert q_2\rangle$ and $\vert q_n\rangle$ is degenerate, as are the pairs $(\vert q_3\rangle,\vert q_{n-1}\rangle)$,
$(\vert q_4\rangle,\vert q_{n-2}\rangle)$, etc. Values $n=2k+1$ will lead to a complete pairing of all states, 
except the vacuum $\vert q_1\rangle$, into degenerate pairs $(\vert q_l\rangle,\vert q_{n-l+2}\rangle)$.
Values $n=2k$ will lead to a similar pairing of degenerate states, with the exception of the vacuum $\vert q_1\rangle$ 
and $\vert q_{k+1}\rangle$, which respectively correspond to the minimum and maximum of $-\cos Q$.
 
Once identified the vacuum $\vert q_1\rangle$ when the kinetic term is switched off, 
we switch it back on to obtain the full Hamiltonian (\ref{gomezcabron}). The latter suggests arranging the basis states 
$\vert q_j\rangle$ in the order 
\begin{equation}
\vert q_n\rangle, \quad\vert q_1\rangle,\ldots, \vert q_{n-1}\rangle.
\label{barbonmaricon}
\end{equation}
In this way, although the full Hamiltonian matrix (\ref{gomezcabron}) is $n\times n$,
it contains a $3\times 3$ submatrix that accounts for the only nonzero
matrix elements involving the state $\vert q_1\rangle$. With the ordering
(\ref{barbonmaricon}), this submatrix lies at the upper, left--hand corner
of the matrix (\ref{gomezcabron}). Explicitly it reads
\begin{equation}
\left(\begin{array}{ccc}
-\cos q_n & -{\rm i}/2 & 0 \\
{\rm i}/2 & -\cos q_1 & -{\rm i}/2 \\
0 & {\rm i}/2 & -\cos q_2
\end{array}\right),
\label{barbonputon}
\end{equation}
with $\cos q_n = \cos q_2$. One can easily verify that the eigenvalues of (\ref{barbonputon}) 
are nondegenerate; its minimum corresponds to a certain $\vert \tilde 0\rangle=a\vert q_n\rangle + b\vert q_1\rangle + 
c \vert q_2\rangle$. For the vacuum of eqn. (\ref{gomezcabron}) we have $\vert 0\rangle=\sum_{j=1}^n c_j\vert q_j\rangle$ 
for some coefficients $c_j$, in general different from the above. However, a glance at $H$ in eqn. (\ref{gomezcabron}) convinces one 
that $\vert \tilde 0\rangle$ tends to $\vert 0\rangle$ as $n$ grows large, while at the same time remaining 
nondegenerate, for the following reasons. 

The addition of the kinetic term $-\cos P$ to the potential $-\cos Q$ might, in principle, force the vacuum out of the 3--dimensional 
subspace of eqn. (\ref{barbonputon}). However this is not the case.
The characteristic polynomial $s_n(E)$ in eqns. (\ref{genial}), (\ref{esplendido}) {\it almost}\/ factorises as 
$s_n(E)=s_3(E)\times s_{n-3}(E)$, the latter being the characteristic polynomial obtained from eqn. (\ref{gomezcabron}) upon removal 
of its upper, left--hand corner (\ref{barbonputon}), plus the boxes $B$, $B^+$ below and to its right, of sizes $(n-3)\times 3$ 
and $3\times(n-3)$ respectively. These boxes are everywhere zero, except for $\langle q_{n-1}\vert B\vert q_n\rangle = -{\rm i}/2 =
-\langle q_3\vert B\vert q_2\rangle$ in $B$ and their complex conjugates in $B^+$. These 4 nonzero matrix elements prevent the factorisation $s_n(E)=s_3(E)\times s_{n-3}(E)$ from being exact. However, their contribution to the characteristic polynomial becomes negligible as $n$ grows large, and the factorisation tends to be exact as $n\to\infty$. If $E_0$ is the vacuum energy of eqn. (\ref{barbonputon}), {\it i.e.}, a root of $s_3(E)$, factorisation ensures that $E_0$ is also a root of $s_n(E)$.

Now the vacuum might develop a degeneracy after the addition of $-\cos P$, {\it i.e.}, $E_0$ might become a multiple root
of $s_n(E)$. However the kinetic term $-\cos P$ acts uniformly on all states $\vert q_j\rangle$, regardless of the value of $j$. 
In the matrix (\ref{gomezcabron}), this kinetic term is responsible for a set of entries all equal to $-{\rm i}/2$ above the diagonal, 
plus a set of entries all equal to ${\rm i}/2$ below the diagonal. This uniformity ensures that the extra energy contributed by 
the kinetic term spreads uniformly over all states, thus respecting their original hierarchy. In particular, the vacuum remains 
nondegenerate.

These arguments establish that, at least in the limit of large $n$, the vacuum of eqn. (\ref{gomezcabron}) is nondegenerate. 
This conclusion is easily seen to hold also when $\beta\neq 1$, $\delta\neq 0$.

\subsection{Dualities}\label{barboncabron}

Next we would like to establish a correspondence between our previous results and the analysis of refs. \cite{MODP, PQM}. 
The vacuum state being nondegenerate, it spans a complex 1--dimensional linear space. Previous sections have determined this linear space on the particular coordinate chart $-\pi<q<\pi$ on the circle $S^1(q)$, where $q$ makes reference to our use of coordinate representation in section \ref{labacasposo}. We can consider a set of transition functions (to be specified presently) so that, under smooth coordinate changes on $S^1(q)$, the vacuum becomes the fibrewise generator of a complex line bundle over the real manifold $S^1(q)$. This line bundle has a real 1--dimensional base and a real 2--dimensional fibre ${\bf C}$. We can manufacture a complex line bundle over the complex torus $T^2=S^1(q)\times S^1(p)$ if we endow the latter with the complex structure of previous sections and pick a set of holomorphic transition functions for the same fibre ${\bf C}$ that existed over $S^1(q)$. In this way the vacuum becomes the fibrewise generator of a complex line bundle over the complex torus.

Now given a complex manifold ${\cal C}$, the elements of its Picard group ${\rm Pic}\ ({\cal C})$ are 1--to--1 with equivalence classes of holomorphic line bundles over ${\cal C}$. The Picard group ${\rm Pic}^{(0)}(T^2)$ of holomorphic line bundles with vanishing first Chern class is another torus, ${\rm Pic}^{(0)}(T^2)=T^2$. The full Picard group ${\rm Pic}\,(T^2)$ consists of an infinite number of copies of ${\rm Pic}^{(0)}(T^2)$, each copy being labelled by an integer $l$ and denoted ${\rm Pic}^{(l)}(T^2)$. 
Thus physically inequivalent vacua over $T^2$ are parametrised by a discrete variable $l\in {\bf Z}$ and by a continuous variable $\lambda\in T^2$. Picking a class $(l, \lambda)\in {\rm Pic}\,(T^2)$ we determine an equivalence class $N(l, \lambda; T^2)$ of holomorphic line bundles over $T^2$ whose fibrewise generator is the vacuum state $\vert 0(l, \lambda)\rangle$. 
In particular, the choice of a class $(l, \lambda)\in {\rm Pic}\,(T^2)$ carries with it the specification of a set of holomorphic transition functions (which were left temporarily undetermined above).

The Picard class $(l,\lambda)$ of the vacuum state of section \ref{pajarescabron} equals $(n,w)$. 
Here we have identified the variable $\lambda\in T^2$ with the complex coordinate 
$w=(\beta q+\delta) +{\rm i}p$. We have also set $l=n={\rm dim}\,{\cal H}$ equal to a positive Chern class; negative
values of $l$ are identified, following ref. \cite{PQM}, with the dimension of the dual space ${\cal H}^*$. 
 
Therefore, the quantum dynamics of (\ref{gomezcasposo}) contains all the elements of ref.  \cite{MODP} 
(concerning the different possible vacua) that are required to implement quantum--mechanical dualities. 
Hence the conclusions of ref.  \cite{MODP} apply to our case, which thus provides an explicit example 
of a quantum--mechanical model exhibiting dualities. In particular, any nonbiholomorphic coordinate transformation 
will be a duality transformation of the quantum theory. As an example, the transformation (\ref{gatita}) is canonical for $\beta =1$ and any $\delta\in(0,2\pi)$, while it is nonholomorphic as different values of $\delta$ generically specify different complex structures. We conclude that the quantum theory is sensitive to a parameter that was irrelevant for the classical theory: varying $\delta$ we have distinct quantum--mechanical models corresponding to a given classical mechanics.

\section{Further examples}\label{alvarezgaumequetepartaunrayo}

Having analysed the torus in detail, we can now easily manufacture examples of higher--dimensional classical phase spaces that exhibit dualities. There are several possibilities. One could consider the quotient of a torus under some group action. Alternatively, one could consider the product of a torus with some other manifold or, more generally, the fibration of a torus over another base manifold. For brevity we will merely give a flavour of these possibilities, concentrating on the specific example of K3 manifolds.

A K3 manifold is a compact, complex surface with the Hodge number $h^{1,0}=0$ and a trivial canonical bundle \cite{JOYCE}.
The Picard group ${\rm Pic}^{(0)}({\rm K3})$ of holomorphic line bundles with vanishing first Chern class is trivial. The full ${\rm Pic}\,({\rm K3})$ consists of an infinite number of copies of ${\rm Pic}^{(0)}({\rm K3})$, each copy being labelled by an integer $l$ and denoted ${\rm Pic}^{(l)}({\rm K3})$. So fixing a degree $l\in{\bf Z}$ we have a unique equivalence class of holomorphic line bundles over K3, and we can certainly vary the vacuum. There is a 20--dimensional space of complex moduli ${\cal M}({\rm K3})$. Although all K3 surfaces are real--diffeomorphic, there are different complex realisations of K3 \cite{JOYCE}: K3 as an orbifold of $T^4$, K3 as a complex surface within ${\bf CP}^3$, and K3 as an elliptically--fibred manifold. 

Take first a 4--dimensional torus $T^4$ parametrised by 2 complex coordinates $z^k$, $k=1,2$. Next consider the ${\bf Z}_2\times{\bf Z}_2$ action on $T^4$ defined by $z^k\rightarrow -z^k$. This action has 16 fixed points, so the quotient space $T^4/({\bf Z}_2\times{\bf Z}_2)$ is singular. Blowing up each one of these singularities with one copy of ${\bf CP}^1$ we obtain the {\it Kummer construction}\/ of K3.
An arbitrary function on $T^4={\bf C}^2/({\bf Z}\times {\bf Z})$ can be expanded as a Fourier series in $\sin q^k$, $\cos q^k$, $\sin p^k$ and $\cos p^k$. On ${\bf C}^2$ with coordinates $z^k=q^k+{\rm i}p^k$, the ${\bf Z}_2\times{\bf Z}_2$ involution reverses the signs of $q^k$ and $p^k$, and only even functions of the latter survive the quotient. Odd functions like $\sin q^k$ and $\sin p^k$, while allowed on $T^4$, would be projected out of K3. Therefore the Hamiltonian analysis performed previously, extended to 2 complex dimensions, also holds for the Kummer construction of K3, even if ${\rm Pic}^{(0)}(K3)\neq {\rm Pic}^{(0)}(T^2)$. For every choice of a vacuum state on Kummer's construction of K3, the previous discussion concerning the variation of the complex structure (while keeping the symplectic structure fixed) remains valid. This leads to dualities for the quantum mechanics defined on the Kummer model of K3.

Next consider, in homogeneous coordinates on ${\bf CP}^3$, the {\it Fermat quartic}
\begin{equation}
(X)^4 + (Y)^4 + (Z)^4 + (T)^4=0,
\label{barbonchupamelapollamarikon}
\end{equation}
which defines an algebraic K3 surface. We can add 19 inequivalent quartic terms to the right--hand side of (\ref{barbonchupamelapollamarikon}), thus giving us a 19--dimensional complex subspace of ${\cal M}({\rm K3})$. Although there are no complex tori involved here, the restriction of the K\"ahler potential on ${\bf CP}^3$ to the quartic defines a dynamics on K3. Thus the quantum--mechanical dualities discussed in ref. \cite{PQM} for projective space also hold on those K3 surfaces given by Fermat quartics.

Finally, a certain family of K3 surfaces admit an elliptic fibration over ${\bf CP}^1$. If the latter has the holomorphic coordinate $w$, 
and $X, Y, Z$ are projective coordiantes on ${\bf CP}^2$, the {\it Weierstrass parametrisation}\/ of the fibration is 
\begin{equation}
Y^2Z=4X^3 - g_2(\Lambda(w)) XZ^2 - g_3(\Lambda(w))Z^3,
\label{cesargomezcasposodemierda} 
\end{equation}
where 
\begin{equation}
g_2(\Lambda(w))=60\sum_{x\in\Lambda(w)-\{0\}}x^{-4},\qquad
g_3(\Lambda(w))=140\sum_{x\in\Lambda(w)-\{0\}}x^{-6}
\label{kasposostodos}
\end{equation}
and $\Lambda(w)$ is the lattice defining the elliptic fibre; its complex structure varies holomorphically as a function of $w$. Over 24 points on the base ${\bf CP}^1$ the elliptic fibre degenerates. Away from those points, given that (locally) such K3's always appear as the product of a torus times a sphere, we can always write down (locally) a dynamics whose Hamiltonian is the sum of two independent Hamiltonians: one for the torus and another one for the sphere as in ref. \cite{PQM}. The complex dimension of ${\cal M}({\rm K3})$ is 18. In this way the dualities discussed on the torus extend to dualities on elliptically--fibred K3 surfaces.

\section{Conclusions}\label{muchasoberbia}

Any function on $T^2$ can be expanded as a Fourier series in $\cos p$, $\sin p$, $\cos q$ and $\sin q$. 
In particular, any Hamiltonian function $H$ on the torus admits such an expansion.
We have treated the case in which $H=-\cos p-\cos (\beta q+\delta)$; the free parameters $\beta$, $\delta$ 
determine a complex structure on $T^2$. We have exhibited the explicit dependence of the quantum theory on the choice 
of a complex structure. Although the classical mechanics described in section \ref{labastidacabron} is insensitive 
to the complex structure, the corresponding quantum mechanics is highly sensitive to it: the Hilbert space of states, 
the operators acting on it, the energy levels, the vacuum state, all depend crucially on the choice of a complex structure 
on classical phase space. Our quantisation being canonical, manifest analyticity (as, {\it e.g.}, in coherent--state 
quantisation) is lost. The dependence of the quantum theory on the complex structure 
appears through its dependence on the parameters $\beta$, $\delta$.

We conclude that, in order to quantise a given classical mechanics, a knowledge of classical phase space (as a real manifold), 
plus the Lagrangian or the Hamiltonian function, does not suffice: we also need to pick a complex structure.
We may then state that \cite{MODP} {\it quantisation is the choice of a complex structure on classical phase space}.
Any two nonbiholomorphic complex structures on classical phase space, even when clasically related by means of a canonical transformation, lead to physically different quantum--mechanical theories. 

We have seen in section \ref{labastidacabron} that the dynamics (\ref{labastidahijoputa}) can be regarded as unifying 
several, apparently different, systems, such as sine--Gordon models and harmonic
oscillators, both of which appear as different approximation regimes of our dynamics (\ref{labastidahijoputa}). 
This situation is very reminiscent of M--theory and the several string theories it unifies.  All differences with M--theory notwithstanding, we may  confidently state that, as suggested in ref. \cite{VAFA}, the relativity of the notion of a quantum is not an exclusive phenomenon 
of fields, strings and branes: as shown here, it also exists within specific quantum--mechanical models.

{\bf Acknowledgements}

It is a great pleasure to thank J. A. de Azc\'arraga for encouragement and
support, and U. Bruzzo for technical discussions.
This work has been partially supported by research grant BFM2002--03681 from 
Ministerio de Ciencia y Tecnolog\'{\i}a and EU FEDER funds.

\end{document}